\documentclass[%
 reprint,
superscriptaddress,
 amsmath,amssymb,
 aps,
]{revtex4-2}

\usepackage{amsmath}
\usepackage{graphicx}
\usepackage{stackengine}
\usepackage[pdftex,breaklinks,linkbordercolor={0.8 0.8 1.0},
citebordercolor={0.8 1.0 0.8},urlbordercolor={0.8 1 1}]{hyperref}

\usepackage[draft]{changes}
\definechangesauthor[name={Francesco Papoff}, color=blue]{fp}
\definechangesauthor[name={Mark Anthony Carroll}, color=red]{mc}
\definechangesauthor[name={Giampaolo D'Alessandro}, color=OliveGreen]{gd}
\definechangesauthor[name={Gian Luca Lippi}, color=violet]{gll}
\definechangesauthor[name={Gian-Luca Oppo}, color=orange]{glo}

\begin{document}

\title{Coherence build up and laser thresholds from nanolasers to macroscopic lasers}%

\author{Mark Anthony Carroll}
\affiliation{Department of Physics, University
of Strathclyde,  107 Rottenrow,  Glasgow G4 0NG, UK.}
\author{Giampaolo D'Alessandro}
\affiliation{School of Mathematical Sciences, University of Southampton, Southampton SO17 1BJ, United Kingdom}
\author{Gian Luca Lippi}
\affiliation{Universit\'e C\^ote d'Azur, Institut de Physique de Nice, UMR 7710 CNRS, 1361 Route des Lucioles, 06560 Valbonne, France}
\author{Gian-Luca Oppo}\affiliation{Department of Physics, University
of Strathclyde,  107 Rottenrow,  Glasgow G4 0NG, UK.}
\author{Francesco Papoff}
\email{f.papoff@strath.ac.uk}
\affiliation{Department of Physics, University
of Strathclyde,  107 Rottenrow,  Glasgow G4 0NG, UK.}

\date{\today}

\begin{abstract} 
We detail the derivation of nanolaser models that include coherent and incoherent variables and predict the existence of a laser threshold, irrespective of cavity size and emitter number, for both single- and multi-electron systems.  The growth in photon number in the lasing mode is driven by an increase in correlation between absorption and emission processes, leading to the onset of self-sustained stimulated emission (laser threshold), followed, in turn, by a correlation decrease and ending with the dominance of coherent emission. The first-order coherence $g^{(1)}$ steadily increases, as the pump grows towards the laser threshold value, and reaches unity at or beyond threshold.  The transition toward coherent emission becomes increasingly sharp as the number of emitters and of the coupled electromagnetic cavity modes increase, continuously connecting, in the thermodynamic limit, the physics of nano- and macroscopic lasers at threshoold.  Our predictions are in remarkable agreement with experiments whose first-order coherence measurements have so far been explained only phenomenologically. A consistent evaluation of different threshold indicators provides a tool for a correct interpretation of experimental measurements at the onset of laser action.
\end{abstract}

\maketitle

\section{Introduction}

The rapid advancement in the design and manufacturing of laser resonators over the past few decades has allowed the construction of lasing devices with mode volume $V \propto \lambda^{3}$, where $\lambda$ is the emission wavelength~\cite{Hill2014advances}. Such small devices are far more compact and less energy hungry compared to standard lasers, as lower input power is required to achieve coherent emission.  In addition to nanolasers, microlasers (e.g., $V = a \lambda^{3}$, $2 \lesssim a \lesssim 40$) hold promise for a number of uses spanning multiple research disciplines and industrial applications, such as integrated optical interconnects, sensing and biological probes, to name a few~\cite{ma2019applications}. Photon number squeezing is also expected to naturally emerge before the transition to coherent emission, leading to cw photon fluxes for non-classical applications~\cite{carroll21b}. The complexity of the transition between incoherent and coherent emission in micro- and nanolasers is at the origin of interpretative problems, and gives rise to new opportunities.  The difficulties in threshold identification in nanolasers~\cite{ning2013laser} come from the intrinsic physical properties of the transition in small systems rather than from technical measurement limitations.

As the mode volume of a device decreases, so does the number of electromagnetic cavity modes available for a spontaneous transfer of the energy stored in the medium. In the Cavity Quantum Electrodynamics (CQED) regime this number is significantly reduced. This ``number'' is characterized by the spontaneous emission factor, $\beta$, which quantifies the ratio between the spontaneous emission rate into the lasing mode and the total spontaneous emission rate. For macroscopic devices $\beta \lesssim 10^{-6}$, while the ideal nanolaser limit corresponds to the asymptote $\beta=1$. In other words, $\beta$ is inversely proportional to the systems size.

The laser threshold is typically identified in macroscopic lasers by inspecting the output power as a function of the input power.  The input-output (I-O) curves display a characteristic S-shape on a log-log plot with a steep growth. The laser threshold is located at the inflection point of these curves~\cite{Rice1994}. As the cavity volume decreases, the steep growth is progressively smoothed, leading, in the nanolaser limit of $\beta = 1$, to a straight line. The extrapolation of this linear dependence down to zero pump power has ushered the questionable concept of thresholdless lasers~\cite{yokoyama89,Rice1994}. 

In spite of the equivalence of the intracavity light-matter interaction in macroscopic and microscopic lasers, two different approaches have emerged, each with its own limitations. For macroscopic systems, the well established semi-classical Maxwell-Bloch equations~\cite{Narducci1988} describe coherent emission above the laser threshold by considering the expectation values of the classical coherent field amplitude and the standard medium polarization. The application of classical factorization schemes to expectation values, which describe the light-matter interaction, neglects quantum correlations~\cite{Kira2011semiconductor}, thus limiting the theoretical description to above-threshold coherent emission, with no access to the incoherent regimes below it.  Quantum models for nanolasers, on the other hand, neglect the classical variables associated with coherent emission~\cite{Chow2013,Chow2014, Kreinberg2017, Gies2007} and apply factorization techniques, such as the cluster expansion~\cite{Fricke1996}, keeping only the slowly varying quantum correlations.

In recent papers~\cite{carroll21a,carroll21b} we combined these two approaches by including the slowly-varying quantum correlations as well as the coherent variables into a single Coherent-Incoherent (CI) model, whilst neglecting quantum correlations between electromagnetic field and those medium operators which oscillate on a fast timescale, as done in semi-classical theories~\cite{Narducci1988}. We then used the Linear Stability Analysis (LSA) of the CI model's incoherent solution to calculate analytically the laser threshold for all two-level emitter nanolasers, including  the so-called {\it thresholdless lasers} ($\beta = 1$).

In this paper we examine in detail the derivation and predictions of the CI model and extend it to multi-electron systems. We discover that -- for both single and multi-electron systems -- the critical point identified by the bifurcation analysis is the threshold beyond which stimulated emission becomes self-sustained.  This analysis is corroborated by experimental measurements~\cite{tempel11}. It is important to stress that our approach further contributes to solving the confusion, first identified in~\cite{Rice1994}, reigning around the concept of a \textit{thresholdless laser} by further dispelling the concept that an ideal CQED laser would be a truly thresholdless device. In a future paper we will consider models that retain fast quantum correlations between field and medium operators, including more quantum aspects than those considered here at the price of a significantly larger number of equations. These models confirm the existence of lasing solutions in all nanolasers~\cite{carroll22a,carroll22b}  and predict laser thresholds associated to the establishment of self-sustained stimulated emission that become increasingly close to those calculated in this paper as the number of emitters increases.

This paper is structured as follows. In Section~\ref{ham} we outline the structure of the system Hamiltonian. Section~\ref{cluster} covers the cluster expansion technique needed to close the model equations and presents the derivation of the CI model for single and multi-electron systems. Section~\ref{lsa} presents the Linear Stability Analysis, including a discussion of the conditions required for the existence of instabilities in the system.  Sections~\ref{thrfeat} and~\ref{betadep} detail the effects of emitter number $N$ and cavity ``size'' $\beta$.  Section~\ref{foc} introduces the characterization of coherence and conclusively interprets existing experimental results in the framework of the models here introduced, while
Section~\ref{concl} offers a brief overview of the work and conclusions.

\section{ The structure of the system Hamiltonian}\label{ham}

Our investigation starts with writing the fully quantized Jaynes-Cummings Hamiltonian~\cite{jaynes1963comparison} generalized to describe light-matter interaction between two interacting levels with lasing and non-lasing modes,
\begin{equation}
  H = H_{free} + H_{int}, 
\end{equation}
where $H_{free}$ is the non-interacting part and $H_{int}$ the interacting part of the Hamiltonian, respectively. The non-interacting part of the Hamiltonian is itself made up of contributions from the free electromagnetic field, $H_{E}$, and the free electrons in the quantum dot, $H_{QD}$,
\begin{equation}
  \label{hfr}
  H_{free} = H_{E} + H_{QD}.
\end{equation}
The photon operators of the system Hamiltonian obey the Bosonic commutation relations, and the carrier operators obey  the Fermi anti-commutation relations. The Bosonic operators $b$ and $b^{\dagger}$ correspond to single-particle operators. It can be shown that 2$N$ Fermi operators are formally equivalent to $N$ Bosonic operators under the requirement that the compound Fermi operators  contain equal numbers of creation and annihilation operators~\cite{Kira2011semiconductor}. Examples are the population of the excited state, $c^{\dagger} c$ ,and the standard material polarization, $v^{\dagger} c$. Therefore, we refer to the coherent field operator $b$ and standard polarization $v^{\dagger} c$ as single particle operators and to the photon number operator $b^{\dagger} b$ and photon assisted polarization $b c^{\dagger} v$ as two particle operators. 

The first term in Eq.~(\ref{hfr}) reads
\begin{equation}
  H_{\mathrm{E}} = \hbar \sum_{q} \nu_{q}
  \left ( b_{q}^{\dagger} b_{q} + \frac{1}{2}\right ),
\end{equation}
where $\nu_{q}$ is the frequency of a photon in the $q$-th mode and the quantum mechanical operators $b_{q}, b_{q}^{\dagger}$ annihilate  and create a photon in the $q$-th mode, respectively. The sum over $q$ includes both lasing and non-lasing modes.

The free electron part of the Hamiltonian describes charge carriers in the conduction and valence band states of the $n$-th quantum dot with respective energies $\epsilon_{c,n}$ and $\epsilon_{v,n}$
\begin{equation}
  H_{QD} = \sum_{n}\left (
    \varepsilon_{c,n} c_{n}^{\dagger} c_{n}
    + \varepsilon_{v,n}v_{n}^{\dagger} v_{n} \right ),
\end{equation}
where $c_{n}, c^{\dagger}_{n}$ and $v_{n}, v^{\dagger}_{n}$ are the annihilation and creation operators, respectively, for conduction and valence electrons of the $n$-th quantum dot.

The two-particle light-matter interaction is described by 
\begin{equation}
  \label{eq:1}
  \begin{split}
    H_{int} = - i\hbar \sum_{n,q}
    & \left [
       g_{nq}\left (b_{q} c_{n}^{\dagger} v_{n}
        + b_{q} v_{n}^{\dagger} c_{n} \right ) \right . \\
      & \left .\, - g_{nq}^{*} \left (b_{q}^{\dagger} v_{n}^{\dagger} c_{n}
        + b_{q}^{\dagger} c_{n}^{\dagger} v_{n}\right )
    \right ],
\end{split}
\end{equation}
where $g_{nq}$ is the light-matter coupling strength between a photon in the $q$-th mode and the $n$-th quantum dot. 

In writing the quantum Hamiltonian we have made the standard assumption of neglecting contributions from phonon and Coulomb interactions between charge carriers. This is equivalent to assuming that the quantum dots are operating at cryogenic temperatures ($\approx 4\mathrm{K}$)~\cite{Chow2014}.

One final remark on the structure of the system Hamiltonian concerns the two-particle operators coming from the interaction part of the Hamiltonian: $b_{q}^{\dagger} c_{n}^{\dagger} v_{n}$ and $b_{q} v_{n}^{\dagger} c_{n}$. Quantum mechanically, operators $b_{q}^{\dagger} c_{n}^{\dagger} v_{n}$  describe a process where a photon in mode $q$ is created in conjunction with the excitation of an electron from the valence to the conduction band of the $n$-th quantum dot. 
Operators $b_{q} v_{n}^{\dagger} c_{n}$  describe its symmetric counterpart, where a photon in mode $q$ is absorbed in conjunction with the de-excitation of an electron from the conduction to the valence band of the $n$-th quantum dot. These two processes do not individually conserve energy, even if their sum is conservative, and oscillate at a frequency approximately double that of the laser, thus in the following we eliminate them from the interaction Hamiltonian (Rotating Wave Approximation).

\section{Cluster Expansion and Nonlinear QED Models}\label{cluster}
\label{Models}

To derive the model equations we work in the interaction picture to obtain Heisenberg's equations of motion for the operators appearing in the system Hamiltonian. The variables that appear in the CI models are the quantum operator expectation values and correlations. The dynamics of an $M$ particle expectation value is directly coupled to an $M+1$ expectation value through equations of the form  
\begin{equation}
  i\hbar d_{t} \langle  M\rangle
  = Lo[\langle 1 \rangle, \cdots, \langle M\rangle]
  + Hi[\langle M + 1\rangle],
\end{equation}
where $d_{t}$ is the first order derivative with respect to time, $\langle 1 \rangle, \cdots, \langle M + 1 \rangle$ indicate the sets of the $1, \cdots, M+1$ particle operators and $Lo$ and $Hi$ are matrices that describe coupling to terms of order $1, \cdots, M$ and $M+1$, respectively. As a result, there is an infinite hierarchy in which each order $M$ depends on the higher order $M+1$.  We must therefore find a way of systematically breaking the infinite hierarchy to obtain a closed set of solvable equations at any order $M$. This is achieved through expressing the expectation values in terms of all possible combinations of products of correlations of lower order operators and introducing approximation schemes for the correlations to truncate the infinite hierarchy~\cite{Kira2011semiconductor}.

The expectation values of photon number and photon-assisted polarization, central to this work, are
\begin{align}
  & \langle b^{\dagger} b \rangle =
    \delta \langle b^{\dagger} b \rangle
    + \langle b^{\dagger} \rangle \langle b \rangle,
    \label{Bb_cl_ex}\\
  & \langle bc^{\dagger} v \rangle =
    \delta \langle b c^{\dagger} v \rangle
    + \langle b \rangle \langle c^{\dagger} v \rangle,
    \label{bCv_cl_ex}
\end{align}
where $\delta \langle b^{\dagger} b \rangle$ is the two particle correlation between emission and absorption and $\delta \langle b c^{\dagger} v \rangle$ is the two particle correlation between photon absorption and electron jump from lower to upper energy level.

We find a closed set of equations by including in the model the expectation values $\langle b \rangle$ and $\langle c^{\dagger} v \rangle$. These correspond to the complex amplitudes of the coherent field and of the medium polarization and have been neglected in previous microscopic models~\cite{Gies2007,Kreinberg2017a,Chow2014}. They display fast oscillations with frequencies of the order of that of the laser mode. On the contrary, the two-particle quantum correlations Eqs.~(\ref{Bb_cl_ex}-\ref{bCv_cl_ex}) that appear in the cluster expansion of the Hamiltonian oscillate slowly. We call the former variables ``coherent'' and the latter ``incoherent''. Coherent quantum correlations such as the correlation between population and field, $\langle b c^{\dagger} c \rangle$, are neglected in the same way as they are in standard semi-classical models. Models that include all possible two particle correlations have also been shown to display laser thresholds~\cite{carroll22a} and will be studied in detail in a future paper. We consider all quantum dots identical. This is not a restrictive hypothesis: we have verified numerically that variations in detuning and light-matter coupling strength up to $10\%$ have negligible effects on the system. Thus, we drop the subscript for the Fermi operators and replace the sum over $n$ with $N$. The resulting system of equations is
\begin{subequations}
  \label{SECNQED}
  \begin{align}
    \label{beq}
    d_{t} \langle b\rangle
    =& -(\gamma_{c} + i\nu)\langle b\rangle
      + N g^{*}\langle v^{\dagger} c\rangle\\
    \label{cdagveq}
    d_{t} \langle c^{\dagger} v\rangle
    =& -(\gamma - i\nu_{\epsilon}) \langle c^{\dagger} v \rangle
      +  g^{*}\langle b^{\dagger}\rangle(2\langle  c^{\dagger} c\rangle - 1) \\
    d_{t} \langle c^{\dagger} c\rangle
    =& r(1-\langle c^{\dagger} c\rangle)
      - (\gamma_{nr}+\gamma_{nl}) \langle c^{\dagger} c\rangle \label{Cc} \\
    &- 2\Re \{g(\delta\langle b_{q} c^{\dagger} v\rangle
      + \langle b_{q}\rangle\langle v^{\dagger} c\rangle) \}   \nonumber  \\
    d_{t} \delta\langle b c^{\dagger} v\rangle
    =& -(\gamma_{c} + \gamma - i\Delta\nu )\delta\langle bc^{\dagger} v \rangle
       \label{deltaBCv}\\
     &+ g^{*}\left [ \langle c^{\dagger} c\rangle
       + \delta\langle b^{\dagger} b\rangle
       \left (2\langle c^{\dagger} c\rangle -1\right )
       - \vert \langle c^{\dagger} v\rangle\vert^{2}
       \right ] \nonumber \\
    d_{t} \delta\langle b^{\dagger} b\rangle
    =& -2\gamma_{c}\delta\langle b^{\dagger} b\rangle
      +  2N \Re \left ( g\delta\langle b c^{\dagger} v\rangle \right ) ,
       \label{eq:2}
  \end{align}
\end{subequations}
where $\nu_{\epsilon}$ is the frequency of the inter-band energy, $h \nu_{\epsilon} = \epsilon_{c}-\epsilon_{v}$, $\Delta \nu \equiv \nu_{\epsilon}-\nu$ is the detuning, $\Re(\cdot)$ stands for real part of its argument and a superscript $^{*}$ indicates the complex conjugate. The equations for $\langle b^{\dagger} \rangle$, $\langle v^{\dagger} c\rangle$ and $\langle b^{\dagger} v^{\dagger} c \rangle$ can be obtained from Eqs.~(\ref{SECNQED}) through complex conjugation. The expectation value of the lower-level population has been eliminated using $\langle c^{\dagger} c\rangle + \langle v^{\dagger} v\rangle =1$. The dissipative part of the equations is obtained by considering Lindblad terms describing the coupling to a Markovian heat bath under the constraint that random excitations into the excited state are neglected~\cite{florian13,leymann14a}, a condition which is fulfilled under the assumed cryogenic temperatures.  The cavity decay rate, $\gamma_{c}$, the polarization dephasing rate, $\gamma$, and the non-radiative losses, $\gamma_{nr}$, describe the dissipative channels. The coherent field amplitude, $\langle b\rangle$, and standard polarization, $\langle v^{\dagger} c\rangle$, are analogous to their amplitudes in semi-classical theories. They describe coherent inter-band processes and therefore need to be externally driven to be sustained. In terms of operators the population density of the excited state, $\langle c^{\dagger} c\rangle$, describes an intra-band process and does not require any externally driven source to exist; it is the probability of an electron being in the excited state. The photon-assisted polarization describes a correlated event between the annihilation of a photon with an inter-band transition or the opposite scenario for its hermitian conjugate. Finally, the intensity correlation describes the correlation between photon absorption and emission.

The model derived above assumes identical emitters, each with two discrete energy levels and a single electron. We want to show that the existence of coherent laser solutions is not specific to the single-electron nature of this model, thus we include the coherent variables $\langle b \rangle$ and $\langle v^{\dagger} c \rangle$ in the model given in Ref.~\cite{Kreinberg2017a}, where the authors relax the single-electron assumption.  To ensure that radiative decays can take place only if the upper level is occupied and the lower level empty, the radiative decay terms are now proportional to the product of the probability that an electron is in the excited level with the probability that the lower level is empty. This gives rise to nonlinear terms in the equation for the photon-assisted polarization and the population density, Eqs.~(\ref{Cc},\ref{deltaBCv}), which in the multi-electron model read
\begin{subequations}
  \label{MECNQED}
  \begin{align}
    d_{t} \langle c^{\dagger} c\rangle
    =& r(1-\langle c^{\dagger} c\rangle)
      - ( \gamma_{nr} + \gamma_{nl} \langle c^{\dagger} c\rangle )
      \langle c^{\dagger} c\rangle\nonumber \\
    &- 2\Re \{g(\delta\langle b_{q} c^{\dagger} v\rangle
      + \langle b_{q}\rangle\langle v^{\dagger} c\rangle) \}
      \tag{\ref{MECNQED}c}\label{Cc-multilevel} \\
    d_{t} \delta\langle b c^{\dagger} v\rangle
    =&  -(\gamma_{c} + \gamma - i\Delta\nu )
      \delta\langle bc^{\dagger} v \rangle
      + g^{*}\left [\langle c^{\dagger} c\rangle^{2} \right . \nonumber \\
    & \left . + \delta\langle b^{\dagger} b\rangle
      ( 2\langle c^{\dagger} c\rangle -1)
      - \vert \langle c^{\dagger} v\rangle\vert^{2}\right ],
      \tag{\ref{MECNQED}d}
  \end{align}
\end{subequations}
while all the other equations remain the same. Note that the population of the lower level, $\langle v^{\dagger} v \rangle$, can be eliminated because $\langle c^{\dagger} c \rangle + \langle v^{\dagger} v \rangle \rightarrow 1$ exponentially in time (see difference between electron and holes densities in Ref.~\cite[Eq.~(3)]{Kreinberg2017a}.

The two sets of equations,~\eqref{SECNQED} and~\eqref{MECNQED}, constitute the single-electron and the multi-electron CI models respectively. They both contain coherent and incoherent variables and differ only in the number of electrons in each emitter. With a quantum theory containing variables that can describe both coherent and incoherent processes, we can now investigate how coherence emerges in nanolasers for single- and multi-electron systems.  

\section{Linear Stability Analysis}\label{lsa}

In analogy with the semi-classical theory of macroscopic lasers~\cite{Narducci1988}, we identify the laser threshold as the instability threshold of a non lasing solution where the incoherent variables are different from zero but the amplitudes of the coherent field are $\langle b\rangle=\langle v^{\dagger} c\rangle=0$.  If perturbations of $\langle b\rangle, \langle v^{\dagger} c\rangle$ grow, the non-lasing solution is unstable.  If instead they decay asymptotically to zero, then the solution is stable. The bifurcation point, where the solution with zero values for coherent variables becomes unstable, is the laser threshold. To study its existence as a function of the parameter values, we perform a linear stability analysis of Eqs.~\eqref{SECNQED} and~\eqref{MECNQED}. We collect coherent and incoherent variables into two groups, $\boldsymbol{c} = \{\langle b\rangle, \langle b^{\dagger} \rangle, \langle v^{\dagger} c\rangle, \langle c^{\dagger} v\rangle\}$ and $\boldsymbol{i} =\{\langle c^{\dagger} c\rangle, \delta\langle b^{\dagger} b\rangle, \delta\langle b c^{\dagger} v\rangle, \delta\langle b^{\dagger} v^{\dagger} c\rangle \}$, respectively, and write the two CI models in a more compact form
\begin{align}
  d_{t} \boldsymbol{i} &= \boldsymbol{F}(\boldsymbol{i}, \boldsymbol{c}),\\
  d_{t} \boldsymbol{c} &= \boldsymbol{G}(\boldsymbol{i}, \boldsymbol{c}),
\end{align}
where $\boldsymbol{G}(\boldsymbol{i}, \boldsymbol{c})$ and $\boldsymbol{F}(\boldsymbol{i}, \boldsymbol{c})$ are non-linear vector functions of $\boldsymbol{i}$ and $\boldsymbol{c}$ whose components are the right-hand side of the two models. With this notation $G_{\langle b\rangle}({\boldsymbol{i}},{\boldsymbol{c}})$ and $G_{\langle v^{\dagger} c\rangle}(\boldsymbol{\boldsymbol{i}},{\boldsymbol{c}})$ are, for example, the right-hand side of Eq.~\eqref{beq} and of the complex conjugate of Eq.~\eqref{cdagveq}, respectively, evaluated at ${\boldsymbol{i}},{\boldsymbol{c}}$. The linearized dynamics of small perturbations $(\boldsymbol{\eta}_{\boldsymbol{i}}, \boldsymbol{\eta}_{\boldsymbol{c}})$ of a fixed point solution $(\boldsymbol{i},\boldsymbol{c})$ is given by
\begin{equation}
  \label{gen_ls}
  d_{t} \left[
    \begin{array}{c}
      \boldsymbol{\eta}_{i}  \\ \boldsymbol{\eta}_{c}
    \end{array}
  \right] = \left[
    \begin{array}{cc}
      \boldsymbol{\nabla}_{\boldsymbol{i}}
      \otimes \boldsymbol{F}({\boldsymbol{i}},{\boldsymbol{c}})
      & \boldsymbol{\nabla}_{\boldsymbol{c}} \otimes
        \boldsymbol{F}({\boldsymbol{i}},{\boldsymbol{c}}) \\ 
      \boldsymbol{\nabla}_{\boldsymbol{i}} \otimes
      \boldsymbol{G}({\boldsymbol{i}},{\boldsymbol{c}})
      & \boldsymbol{\nabla}_{\boldsymbol{c}} \otimes
        \boldsymbol{G}({\boldsymbol{i}},{\boldsymbol{c}})
    \end{array}
  \right] \left[
    \begin{array}{c}
      \boldsymbol{\eta}_{\boldsymbol{i}} \\
      \boldsymbol{\eta}_{\boldsymbol{c}}
    \end{array} \right] .
\end{equation}
Each block in the matrix on the right-hand side of Eq.~\eqref{gen_ls} is of dimension $4 \times 4$ (both sets ${\boldsymbol{i}}$ and ${\boldsymbol{c}}$ contain four variables) and corresponds to the Jacobian with respect to the $\boldsymbol{i}$ and $\boldsymbol{c}$ variables.  $\otimes$ denotes the outer product. For any solution with ${\boldsymbol{c}}=\boldsymbol{0}$ one has $\boldsymbol{\nabla}_{\boldsymbol{i}} \otimes \boldsymbol{G}({\boldsymbol{i}},\boldsymbol{0})=\boldsymbol{0}$ and $\boldsymbol{\nabla}_{\boldsymbol{c}} \otimes \boldsymbol{F}({\boldsymbol{i}},\boldsymbol{0})=\boldsymbol{0}$, so that coherent and incoherent perturbations of the $({\boldsymbol{i}},\boldsymbol{0})$ solutions decouple.

This is a general feature of \emph{all} models derived under the rotating wave approximation independently of the order of the quantum correlations considered.
Its origin is the separation of time scale between the coherent and the incoherent variables. The (fast) coherent variables oscillate at the lasing frequency $\nu$, i.e., proportionally to $\sim e^{-i \nu t}$. They can therefore only appear in complex conjugate quadratic pairs in the equations for the (slow) incoherent variables. As the derivative of a quadratic term at zero is zero, we have that $\boldsymbol{\nabla}_{\boldsymbol{c}} \otimes \boldsymbol{F}({\boldsymbol{i}},\boldsymbol{0})=\boldsymbol{0}$. Conversely, the (slow) incoherent variables can only appear in the equations for the (fast) coherent variables if they are multiplied by a coherent variable.  Therefore, $\boldsymbol{\nabla}_{\boldsymbol{i}} \otimes \boldsymbol{G}({\boldsymbol{i}},\boldsymbol{0})=\boldsymbol{0}$.

While these results are generic, in the specific case of the CI models the incoherent perturbations of the incoherent solution always decay to zero. Therefore, the existence of a laser threshold is determined solely by the dynamics of the coherent perturbations, given by
\begin{equation}
  d_{t} \boldsymbol{\eta}_{\boldsymbol{c}}
  = \boldsymbol{\nabla}_{\boldsymbol{c}} \otimes
  \boldsymbol{G}({\boldsymbol{i}},\boldsymbol{0})
  \boldsymbol{\eta}_{\boldsymbol{c}}
  = \left[
    \begin{array}{cc}
      J & 0 \\ 
      0 & J^{*}
    \end{array}
  \right] \boldsymbol{\eta}_{\boldsymbol{c}}, 
\end{equation}
where
\begin{align}
  J
  &= \left[
    \begin{array}{cc}
      \partial_{\langle b\rangle}
      G_{\langle b\rangle}({\boldsymbol{i}},\boldsymbol{0})
      & \partial_{\langle v^{\dagger} c\rangle}
        G_{\langle b\rangle}({\boldsymbol{i}},\boldsymbol{0})\\ 
      \partial_{\langle b\rangle}
      G_{\langle v^{\dagger} c\rangle}({\boldsymbol{i}},\boldsymbol{0})
      & \partial_{\langle v^{\dagger} c\rangle}
        G_{\langle v^{\dagger} c\rangle}({\boldsymbol{i}},\boldsymbol{0})
    \end{array}
        \right]
  \nonumber \\
  &= \left[
    \begin{array}{cc}
      -\gamma_{c} & g^{*}N \\ 
      g(2\langle c^{\dagger} c\rangle - 1) & -(\gamma + i\Delta\nu)
    \end{array}
  \right],
\end{align}
and $J^{*}$ is the complex conjugate of $J$. For ease of notation and without loss of generality, we have written $J$ in a frame rotating with $\langle b\rangle$. This matrix depends on the system parameters and on the population of the excited state. It is important to note that the structure of the stability matrix is the same for both single- and multi-electron CI models. However, since $J$ depends on the excited state population the eigenvalues of these two models differ.

The lasing threshold condition is that there is at least one eigenvalue $\lambda$ of $J$ such that $\Re(\lambda)>0$.  Since $0\leq \langle c^{\dagger} c\rangle \leq 1$, this can be satisfied only if
\begin{equation}
  N > \frac{\gamma\gamma_{c}}{\vert g\vert^{2}}
  \left [ 1 + \left ( \frac{\Delta\nu}{\gamma+\gamma_{c}} \right )^{2} \right],
  \label{eq:scond}
\end{equation}
i.e., if the number of quantum dots is greater than a critical number given by the right hand side of Eq.~(\ref{eq:scond}). This applies to both the single- and multi-electron models, is independent of $\beta$, and increases with losses and detuning.  We conclude this section with two observations. The first is that the CI models have been derived assuming weak light-matter coupling. Therefore Eq.~(\ref{eq:scond}) does not apply to the strong coupling regime. The second, is that the instability condition on the number of emitters in Eq.~(\ref{eq:scond}) is only a necessary one: a sufficiently large pump rate is also necessary to cross the laser threshold, as discussed in the following.

\begin{figure}
  \centering
  \includegraphics[width=\columnwidth]{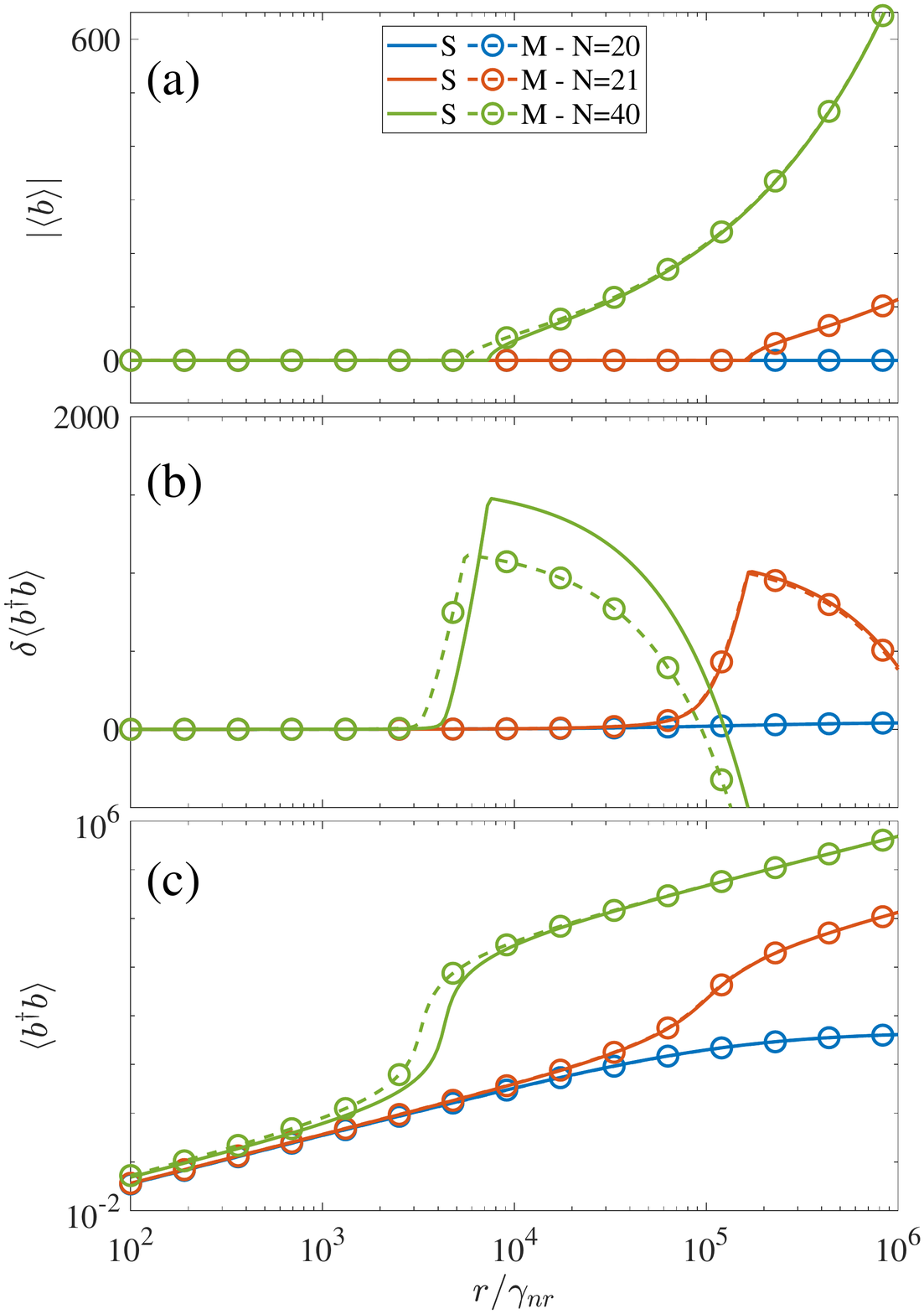}
  \caption{(a) The modulus of the coherent field amplitude $|\langle b \rangle|$, (b) the correlation between photon absorption and emission, $\delta \langle b^{\dagger} b \rangle$, and (c) the expectation value of the photon number, $\langle b^{\dagger} b \rangle$ as a function of the pump for the single- (solid line) and the multi-electron (dashed with circles) CI models, for $N=\{20, 21, 40\}$ (blue, red and green lines respectively). In this and all other figures time and decay and coupling parameters are scaled with $\gamma_{nr}$, which is equivalent to setting $\gamma_{nr}=1$ in Eqs.~\eqref{SECNQED} or~\eqref{MECNQED}. The other parameter values are $g=70$, $\Delta \nu=0$, $\gamma = 10^{4}$,  $\gamma_{c} = 10$ and $\gamma_{nl}=1400$, equivalent to $\beta=7\times 10^{-4}$.}
  \label{fig:fig2}
\end{figure}

\section{Laser threshold:  dependence on N}\label{thrfeat}

We now investigate photon emission processes in these single- and multi-electron lasers below and above the instability threshold. Since $\langle b\rangle=0$ when the device is not lasing, we see from the cluster expansion in Eq.~\eqref{Bb_cl_ex} that the photon number is given exclusively by the correlation term, which dominates the spontaneous emission regime. Fig.~\ref{fig:fig2} a) illustrates the effect of including the fast variables for the single-electron (solid lines) and the multi-electron (dashed lines with circles) CI models, Eqs.~\eqref{SECNQED} and~\eqref{MECNQED}. We compare three devices containing $N=\{20,21,40\}$ emitters (blue, red and green lines respectively). For the parameter values of this illustration (see caption of Fig.~\ref{fig:fig2}), an instability exists if $N>20$. Below the critical number of quantum dots (blue lines), as the pump increases the photon number saturates and the coherent field amplitude remains zero, confirming the absence of laser emission. For a number of quantum dots just above the minimum number required for lasing, i.e., $N=21$ (red line), there is a clear jump in the photon number accompanied by an emerging non-zero coherent field amplitude via a pitchfork bifurcation (Fig.~\ref{fig:fig2}a). We can see from the graph of $\delta\langle b^{\dagger} b\rangle$, Fig.~\ref{fig:fig2}b, that the initial growth in photon number is due to spontaneous emission, positively and increasingly correlated to absorption, while the coherent part of the field is zero. Indeed, the growth of $\delta\langle b^{\dagger} b\rangle$  in Fig.~\ref{fig:fig2}b  precedes the bifurcation (Fig.~\ref{fig:fig2}a) and occurs at pump rates toward the end of the steeper growth in the photon number that is visible in Fig.~\ref{fig:fig2}c. This is a characteristic feature which distinguishes small from macroscopic lasers.  For the latter, it is known that the inflection point of the steeper photon number growth corresponds to the threshold~\cite{Rice1994} and, by extension, this points has been taken as a reference also for small lasers with the help of clever techniques~\cite{ning2013laser}.  Finite-size effects, instead, profoundly modify not only the nature of threshold (as explained below), but also the pump value for which it occurs.  The consequences are important since the identification of coherent emission becomes problematic.  The difficulty is pragmatically circumvented, in commercial microdevices, by manufacturers~\footnote{See, e.g., the Manufacturer's specification on the Thorlabs site for VCSEL980 (microVCSEL for data transmission).} whose laser characteristic sheets give a threshold current which is placed well beyond the actual threshold, identified here through linear stability analysis.  A discussion of the various ``kinds'' of threshold experimentally used is offered in~\cite{lippi22} (Supplementary Material available in~\cite{lippi22ar}).

From threshold onward, the increase of the coherent field intensity $|\langle b^{\dagger} \rangle|^{2}$ coincides with a sharp decrease in the correlation between absorption and emission, $\delta\langle b^{\dagger} b\rangle$, as expected in the presence of stimulated emission. When the correlation $\delta\langle b^{\dagger} b\rangle$ becomes negative, stimulated emission dominates and $ \langle b^{\dagger} b\rangle = |\langle b^{\dagger} \rangle|^{2} + \delta\langle b^{\dagger} b\rangle< |\langle b^{\dagger} \rangle|^{2}$.  In summary, Fig.~\ref{fig:fig2}b clearly shows two features: ($i$) during the steep parts of the emission growth the light is entirely incoherent; ($ii$) immediately above threshold, the emitted field consists of a mixture of coherent and incoherent photons and complete dominance of the coherent component takes place only (well) beyond threshold.  These features are typical of nano- and microlasers.

\begin{figure}
  \centering
  \includegraphics[width=\columnwidth]{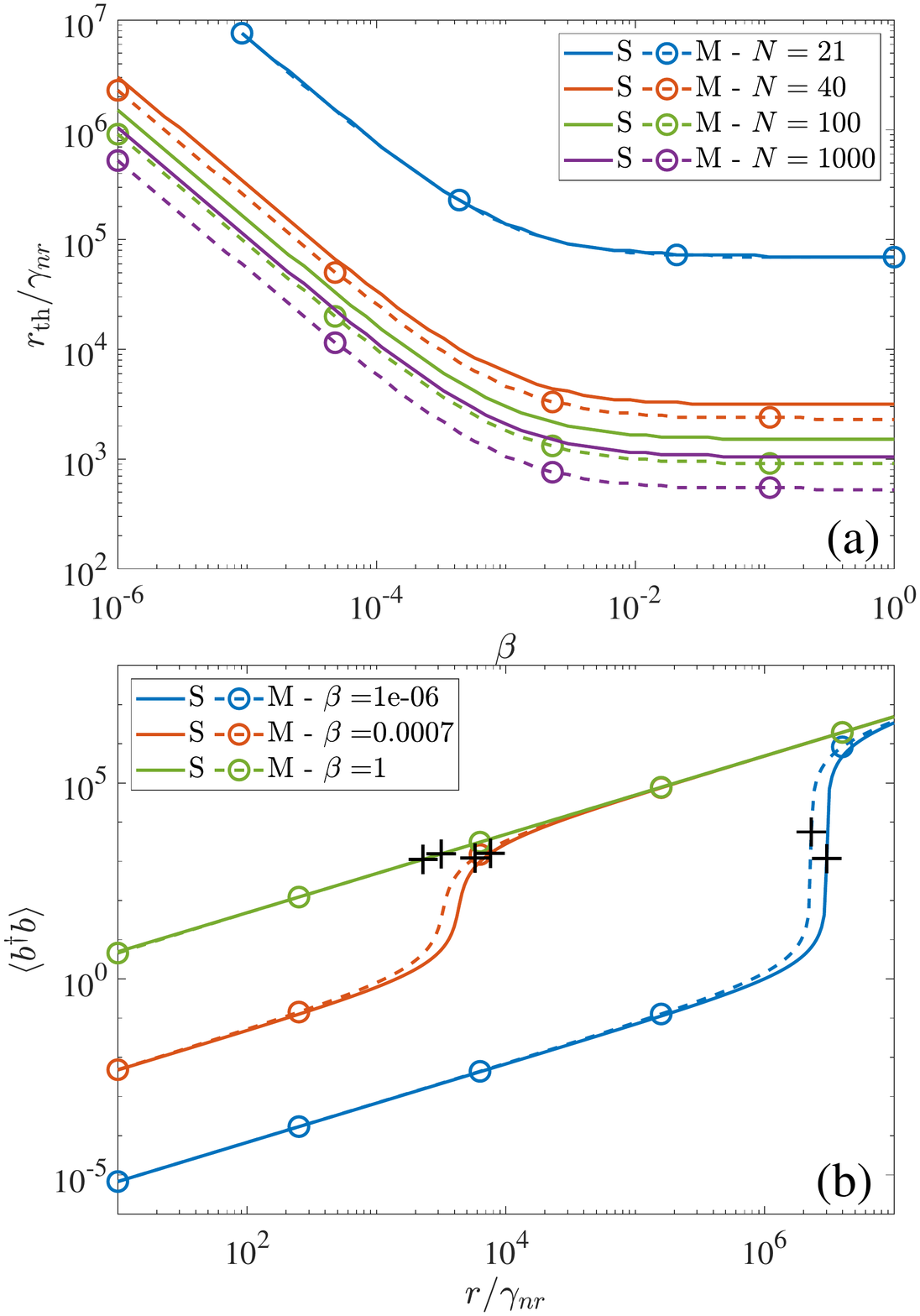}
  \caption{ (a) Numerical estimate of the pump threshold for the single- (S) and multi-electron (M) CI models as a function of the spontaneous emission factor, $\beta$, for different numbers of emitters. (b): Photon number $\langle b^{\dagger} b \rangle$ as a function of the pump $r$ for the single- (S) and multi-electron (M) CI models for different values of $\beta$ and $N=40$ quantum dots . The black crosses identify the numerically established laser thresholds for the two models. All other parameters as in Fig.~\ref{fig:fig2}.}
  \label{fig:fig5}
\end{figure}

The smoothness of the lasing transition imposed by the finite size of the (small) devices paves the way towards new applications~\cite{wang21a,wang21b}. In contrast, the sharpness of the transition in macrolasers (e.g., Fig.~\ref{fig:fig5}) squeezes the pump interval over which the evolution from entirely incoherent to dominantly coherent emission takes place, explaining why, in macroscopic lasers, the threshold can be considered as an on-off effect that corresponds to a single well defined pump value. It is a strength of the CI models that they provide a description of the continuous transformation in the laser emission features as its size increases.

A device with twice the minimum number of quantum dots  (e.g., $N=40$,  dashed lines in Fig.~\ref{fig:fig2}) crosses the laser threshold at a pump rate lower than that for $N=21$ and with a sharper transition, see Fig.~\ref{fig:fig2}c.  
  
As $N$ increases the differences between the single- and multi-electron model become apparent (compare the solid and dashed curves in Fig.~\ref{fig:fig2}). The multi-electron model reaches threshold for lower values of the pump rate and, hence, the fraction of incoherent emission contributing to the initial growth in photon number is reduced. This is due the lower losses of the upper level population, $\langle c^{\dagger} c \rangle $, due to the term $\gamma_{nl} \langle c^{\dagger} c \rangle^{2}$ in Eq.~\eqref{Cc-multilevel} compared to the losses due to the term $\gamma_{nl} \langle c^{\dagger} c \rangle$ in Eq.~\eqref{Cc}. Both models have the same critical number of emitters necessary for the instability to exist ($N=21$ for the chosen parameters). Only the pump power at the laser threshold changes.  

These results highlight the contributions of the fast variables, and the necessity of their presence in the models to obtain a consistent description of the emission processes in a laser. The position of the laser bifurcation in the I-O curve shows that a simple visual inspection of the output characteristics leads to an incorrect identification of the laser threshold and fails to identify the true nature of the emission process, e.g., the incoherent nature of the photon number in small lasers in the phase of steep growth. 

\section{Laser threshold: dependence on $\beta$}\label{betadep}

We now turn to the dependence of threshold on system size, $\beta$~\cite{carroll21a}. Fig.~\ref{fig:fig5}a displays the value of pump at threshold as a function of $\beta$ for devices with different $N$.  This has been computed numerically by finding the pump value for which the correlation $\delta \langle b^{\dagger} b \rangle$ is maximum (see Fig.~\ref{fig:fig2}b). While the threshold pump rate decreases monotonically as $N$ increases (for all values of $\beta$), the dependence on cavity size shows the existence of two regimes: a rapid threshold decrease within the realm of macroscopic lasers, and the onset of near-saturation (in double logarithmic scale) for $\beta \gtrapprox 10^{-3}$ (i.e., micro- and nanolasers).  This latter feature would appear to contradict the common knowledge according to which the threshold linearly decreases with slope $\frac{1}{2}$ in double logarithmic scale~\cite[Eq.~(20)]{Rice1994}) as $\beta$ increases; this property is, however, based exclusively on the (incorrect) assumption that threshold is always placed at the inflection point of the I-O curve.  Instead, the saturation which emerges from the CI models results from the identification of the true laser threshold (self-sustained stimulated emission, Section~\ref{lsa}) which progressively and substantially moves away from the macroscopic definition as the laser size is decreased.  The loss in threshold reduction is, however, well compensated by the emergence of a broader and richer transition region between incoherent and coherent emission, whose features promise new applications (Sections~\ref{lsa} and~\ref{foc}).

A clear visual illustration of the threshold displacement is provided in Fig.~\ref{fig:fig5}b, showing the I-O curves in double logarithmic scale for laser devices with $N = 40$ emitters and three different values of $\beta$.  The straight, superposed I-O curves correspond, as expected, to $\beta = 1$, while those with a gentle curvature to a microlaser: the respective thresholds are marked by black crosses and appear well on the upper branch.  It is only with a macroscopic laser that the threshold appears at the inflection point of the steeply growing photon number, matching the well-known properties of macroscopic lasers~\cite{Rice1994}.

\section{First order correlation function $g^{(1)}(\tau)$}\label{foc}

Surprisingly not included in the recommendations to identify laser threshold in experiments~\cite{lasthr}, the autocorrelation functions remain the most sensitive and most reliable way of obtaining pertinent threshold information, as long as a meaningful model can be used for comparison.  In this section, we tackle precisely this aspect and examine the first order, time-delayed correlation function.  Here we study its properties and successfully compare them to the experimental measurements in Ref.~\cite{tempel11}.  Due to its more complex experimental implementation, it is more seldom used than its second order counterpart, but it has the advantage of providing direct information on the coherence of the emitted radiation~\cite{tempel11}.   Once the relationship between the two kinds of correlations is clarified, comparison between the two indicators will facilitate their individual use in the interpretation of experimental results.

In order to calculate the first-order correlation function
\begin{equation}
  \label{eq:3}
  g^{(1)}(\tau) = \frac{\langle b^{\dagger}(t) b(t+ \tau) \rangle}
  {\langle b^{\dagger}(t) b(t) \rangle} ,
\end{equation}
where $\tau$ is a delay time, we write the differential equation
\begin{equation}
  \label{eq:4}
  d_{\tau} g^{(1)} = \frac{1}{\langle b^{\dagger}(t) b(t) \rangle}
  d _{\tau} \langle b^{\dagger}(t) b(t+\tau) \rangle 
\end{equation}
which we solve with initial condition $g^{(1)}(0)=1$.  To form a close set of equations, we use the quantum regression formula, see Eqs.~(1.105-1.107) of Ref.~\cite{carmichael2002}. In the Heisenberg picture, this reads $d_{\tau} \langle A(t) B(t+\tau) \rangle = \langle A(t)d_{\tau} B(t+\tau) \rangle$ where $A$ and $B$ are operators and $d_{\tau} B$ is calculated applying the Hamiltonian and Lindblad formalism at time $t+\tau$. We expand the $\tau$ derivative on the right hand side of Eq.~\eqref{eq:4} and make use of Eqs.~\eqref{beq} and~\eqref{deltaBCv} to obtain
\begin{subequations}
  \label{newg0}
  \begin{align}
    & d_{\tau} \langle \tilde{b}^{\dagger} b\rangle
      = -(\gamma_{c} + i\nu)\langle \tilde{b}^{\dagger} b\rangle
      + N g^{*}\langle \tilde{b}^{\dagger} v^{\dagger} c\rangle,\\
    &d_{\tau} \langle \tilde{b}^{\dagger} v^{\dagger} c\rangle
      = -(\gamma + i\nu_{\epsilon})
      \langle \tilde{b}^{\dagger} v^{\dagger} c \rangle
      +  g \left (
      2 \langle  \tilde{b}^{\dagger} b c^{\dagger} c \rangle
      - \langle \tilde{b}^{\dagger} b \rangle
      \right ),
  \end{align}
\end{subequations}
where $ \tilde{b}^{\dagger} \equiv b^{\dagger}(t) $, and all other operators are at time $t+\tau$. $\langle  \tilde{b}^{\dagger} b c^{\dagger} c \rangle$ is the expectation value of a $3$-particle operator. To find a closed set of equations at two-particle level we use Eq.~(\ref{Bb_cl_ex}) and the cluster expansion
\begin{equation}
  \begin{split}
    \langle  \tilde{b}^{\dagger} b c^{\dagger} c \rangle
    =& \delta \langle  \tilde{b}^{\dagger} b c^{\dagger} c \rangle
    + \langle \tilde{b}^{\dagger} \rangle 
    \delta \langle b c^{\dagger} c \rangle \nonumber \\
    & + \langle b \rangle
    \delta \langle \tilde{b}^{\dagger} c^{\dagger} c \rangle
    + \langle c^{\dagger} c \rangle
    \delta \langle \tilde{b}^{\dagger} b\rangle
    + \langle  \tilde{b}^{\dagger} \rangle \langle  b \rangle
    \langle  c^{\dagger} c \rangle
  \end{split}
\end{equation}
together with the semi-classical approximation used to derive the CI models, which for these equations reduces to $\delta \langle \tilde{b}^{\dagger} c^{\dagger} c \rangle \sim \delta \langle b c^{\dagger} c \rangle \sim 0$. With these approximations Eqs.~(\ref{newg0}) become
\begin{subequations}
  \label{g1}
  \begin{align}
    & d_{\tau} \langle \tilde{b}^{\dagger} b\rangle
      = -(\gamma_{c} + i\nu)\langle \tilde{b}^{\dagger} b\rangle
      + N g^{*}\langle \tilde{b}^{\dagger} v^{\dagger} c\rangle,\\
    & d_{\tau} \langle \tilde{b}^{\dagger} v^{\dagger} c\rangle
      = -(\gamma + i\nu_{\epsilon})
      \langle \tilde{b}^{\dagger} v^{\dagger} c \rangle
      + g \langle \tilde{b}^{\dagger} b\rangle
      (2\langle  c^{\dagger} c\rangle - 1).
  \end{align}
\end{subequations}
Eqs.~(\ref{g1}) are formally identical for the single- and multi-electron models, the only difference in $g^{(1)}(\tau)$ coming from the different values of the term $\langle  c^{\dagger} c\rangle$. This is due to the fact that the Heisenberg equations and the dissipative Lindblad terms for the operators at time $t+\tau$ do not depend on the losses of $\langle c^{\dagger} c \rangle$ that are proportional to $\gamma_{nl}$. 

\begin{figure}
  \centering
  \includegraphics[width=\columnwidth]{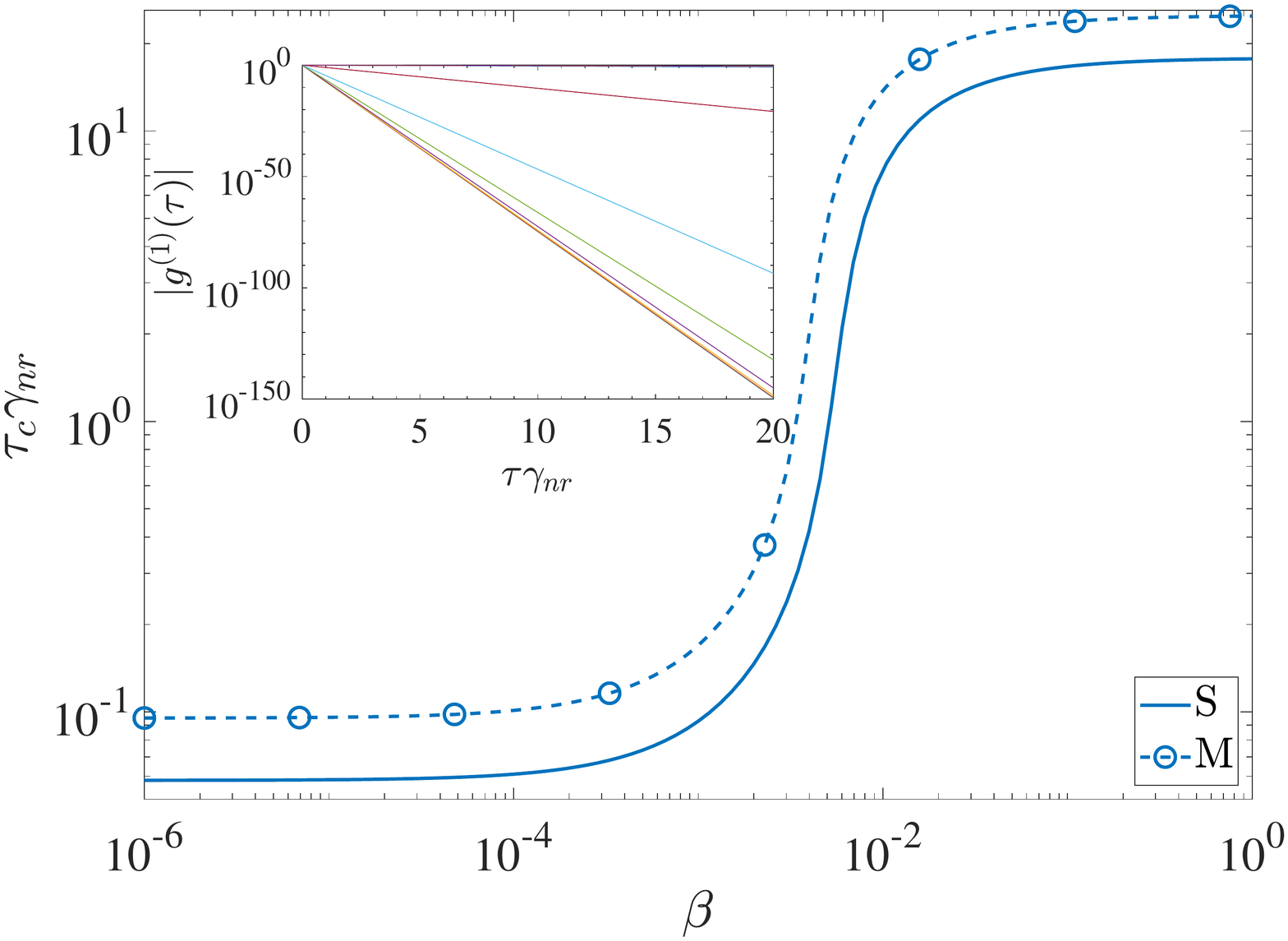}
  \caption{The main graph plots the coherence decay time $\tau_{c}$ as a function of the spontaneous emission factor $\beta$ for the single- (S) and multi-electron (M) CI models for a pump value equal to 15\% of the threshold for $N=40$ quantum dots. $\tau_{c}$ has been obtained by fitting with a straight line $\log(|g^{(1)}(\tau)|)$ as a function of $\tau$. The inset is a log plot $|g^{(1)}(\tau)|$ as a function of the delay time $\tau$ for a sample of the values of $\beta$ used in the main plot. This confirms the exponential decay of the correlation, Eq.~\eqref{eq:5}.  All other parameters as in Fig.~\ref{fig:fig2}.}
  \label{fig:fig3}
\end{figure}

With the help of these expressions, we can now plot the first order autocorrelation as a function of the model parameters. We expect that below threshold the correlation function decays exponentially with the delay time,
\begin{equation}
  \label{eq:5}
  g^{(1)}(\tau) \propto e^{-t/\tau_{c}},
\end{equation}
with $\tau_{c}$ the correlation decay time.  This behavior is confirmed by the log plot of $g^{(1)}(\tau)$ in the inset of Fig.~\ref{fig:fig3}, where we have set the pump at 15\% of the single-electron threshold value~\cite[Eq.~(20)]{carroll21c} for $N=40$ quantum dots.  We have computed $\tau_{c}$ as a function of $\beta$ by fitting these curves with a straight line. The decay rate has a sigmoidal behavior: it is an increasing function of $\beta$ that jumps by two orders of magnitude as $\beta$ changes from $10^{-4}$ to $10^{-2}$ and is approximately constant outside this interval. This clearly illustrates a fundamental feature of small lasers, whose coherence grows gradually as threshold is approached, in agreement with the smooth response of their I-O curve.  For macroscopic lasers, on the other hand, we obtain results which are consistent with the standard picture of a nearly incoherent output up until threshold, with a sudden conversion to full coherence. The single- and multi-electron CI models have similar behavior, with the multi-electron model having larger decay time. This is an effect of the lower effective losses of the multi-electron with respect to the single-electron model: at equal pump values the former is closer to threshold than the latter (cf. the shift of threshold positions between the two models in Fig.~~\ref{fig:fig5}).

\begin{figure}
  \centering
  \includegraphics[width=\columnwidth]{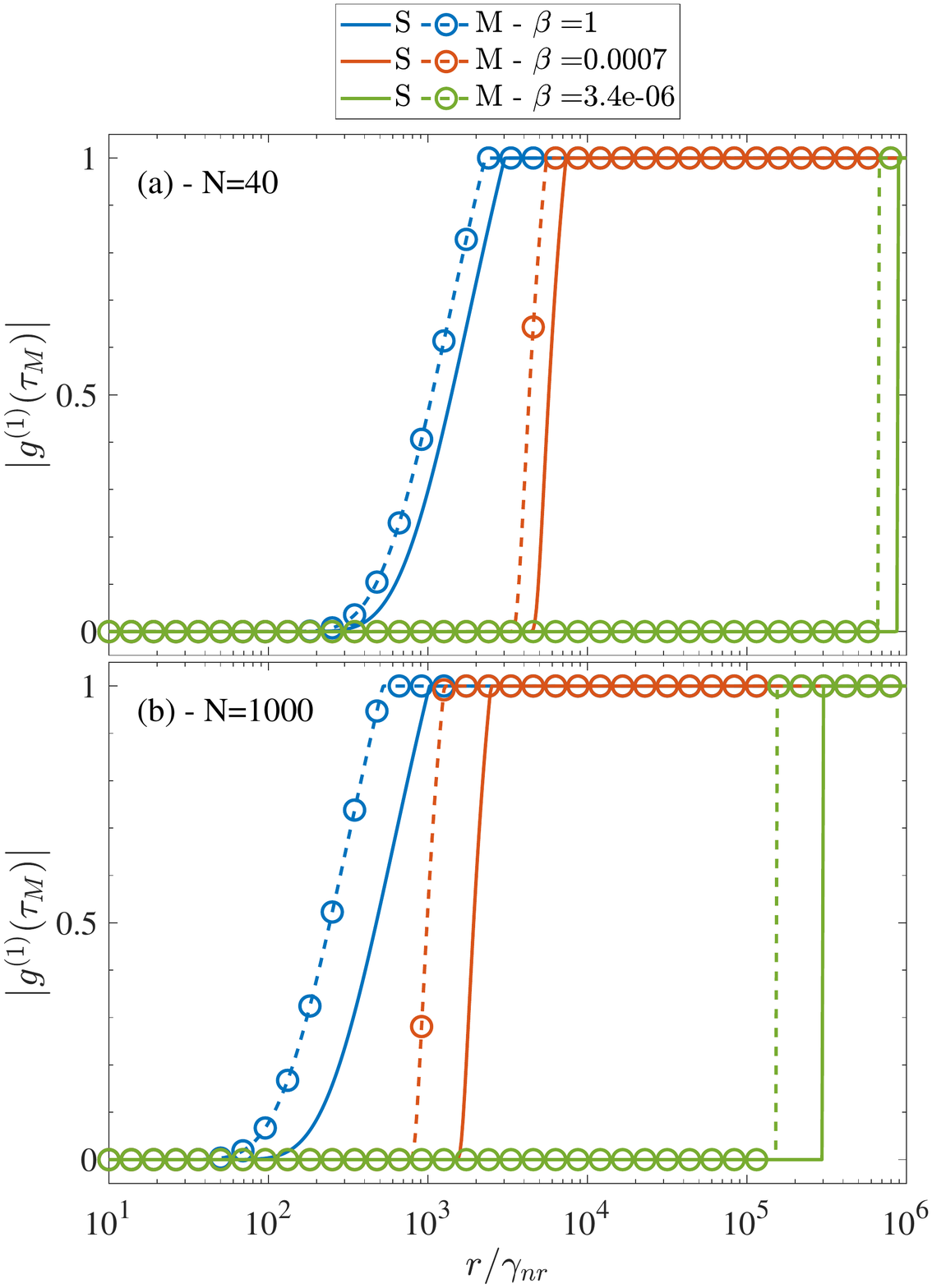}
  \caption{ $\left | g^{(1)} \right |$ for delay $\tau_{M} = 60 / \gamma_{nr}$ as a function of the pump for $\beta = \{1, 7 \times 10^{-4}, 3.4 \times 10^{-6}\}$ (blue, red and green lines respectively) for the single- (solid) and multi- (dashed with circle) CI models. The number of quantum dots is $N=\{40,1000\}$ in panels (a) and (b) respectively. The points where $g^{(1)}(\tau_{M})$ reaches unity, and where the slopes of the curves suddenly change, are the laser thresholds.  All other parameters as in Fig.~\ref{fig:fig2}.}
    \label{fig:fig7}
\end{figure}

The evolution of coherence with pump power is examined in Fig.~\ref{fig:fig7} for the single- and multi-electron models at fixed number of emitters, $N = \{40,1000\}$, and cavity volumes, $\beta = \{1, 7 \times 10^{-4}, 3.4 \times 10^{-6}\}$.  In order to clearly highlight the pump influence, a delay time $\tau = 60/\gamma_{nr}$ is fixed.  Experimental information can be gathered, as in~\cite{tempel11}, by fixing the difference in the Michelson interferometer arm lengths and measuring the fringe visibility as a function of pump.  The laser threshold corresponds to the smallest pump value for which $g^{(1)}(\tau_{M}) = 1$; at this point the curve slope is discontinuous. While in an experiments unavoidable fluctuations of the control parameters, not included in the model, will limit the coherence time even beyond the threshold, the behavior of the coherence time as a a function of the pump will provide a clear indication of the threshold value. Irrespective of laser size, there is a continuous growth of coherence, driven by the increase in correlation between absorption and emission properties (as in Fig.~\ref{fig:fig5}); however, while in smaller systems coherence evolves steadily over a broad pump range below threshold, in macroscopic lasers the change occurs over a narrow interval of pump values.  In other words, as $\beta$ decreases moving toward the macroscopic limit, it becomes more and more difficult to obtain partially coherent emission.  This result does not depend on the choice of $\tau_{M}$, as long as $\tau_{M} \gg \lambda_{0}/v$, with $\lambda_{0}$ and $v$ the light wavelength and velocity in the interferometer, respectively. Changing $\tau_{M}$ only changes the shape of the curves of Fig.~\ref{fig:fig7}.  It is worth stressing again that the deformation of the coherence curves progresses continuously from the nano- to the macroscale.

Increasing the number of emitters from $N=40$, Fig.~\ref{fig:fig7}a, to $N = 10^{3}$, Fig.~\ref{fig:fig7}b, while keeping the other parameters constant reduces the threshold values and the range of pump values over which the transition toward $g^{(1)}(\tau_{M}) = 1$ occurs for all values of $\beta$.  While nanolasers are typically built with tens of emitters, in macroscopic lasers their number will easily be largely in excess of what we are showing here, thus further enhancing the differences between the two categories of devices.

\begin{figure}
  \centering
  \includegraphics[width=\columnwidth]{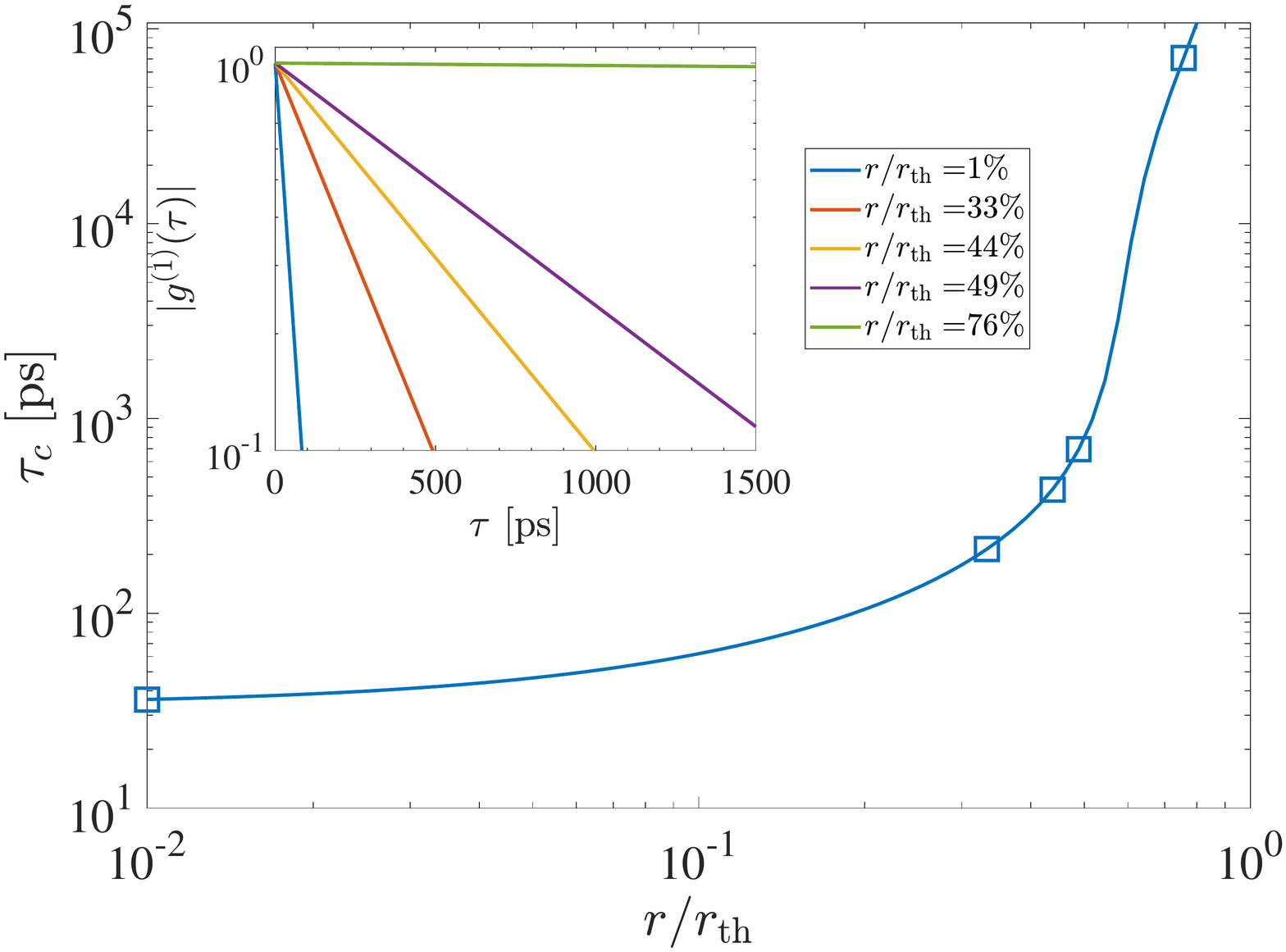}
  \caption{The main graph plots the coherence decay time $\tau_{c}$ as a function of the pump power in units of the single electron CI threshold for $\beta=7 \times 10^{-4}$ and $N=40$ quantum dots. $\tau_{c}$ has been obtained by fitting with a straight line $\log(|g^{(1)}(\tau)|)$ as a function of $\tau$. The inset is a log plot $|g^{(1)}(\tau)|$ as a function of the delay time $\tau$ for the pump values indicated by square symbols in the main plot.  All other parameters as in Fig.~\ref{fig:fig2}. This figure is the analogous of figures~2b and 2c of Ref.~\cite{tempel11}. For ease of comparison time units are expressed in picoseconds. The dimensional time scale has been fixed by setting $\gamma_{nr}=1$~ns.}
\label{fig:fig4}
\end{figure}

We conclude this section by highlighting that these analytical and numerical results are supported by independent experimental measurement of $g^{(1)}(\tau)$~\cite{tempel11}.  The first order coherence was experimentally obtained in Ref.~\cite{tempel11} by measuring the visibility of interference fringes resulting from Michelson interferometry and plotted as a function of the pump power~\cite[Fig.~2b]{tempel11}.  From these data the authors also computed the coherence decay time as a function of the pump power.  It is not possible from the experimental data available in Ref.~\cite{tempel11} to obtain unique values for the CI model parameters.  However, the parameter values used in the figures in this paper are reasonable estimates. We plot in Fig.~\ref{fig:fig4} the correlation decay time as a function of the pump power, measured in units of the analytical threshold for the single electron CI model~\cite[Eq.~(20)]{carroll21c}. The similarities between this figure and its inset and figures 2c and 2b respectively of Ref.~\cite{tempel11} are uncanny, keeping in mind the uncertainty in the mapping of the experimental parameters.  We can therefore conclude that the CI model is capable of clearly and unequivocally identifying the onset of coherence, matching it to the crossing of laser threshold (self-sustained growth of stimulated emission), and to explain experimental observations for which no model, derived from first principles, had been available up until now.

\section{Conclusions}\label{concl}

We have presented the details of a model for (semi-conducting) quantum emitters (with single- or multiple-electrons) coupled to an electromagnetic cavity of arbitrary size to describe the transition from thermal to coherent emission.  The joining of a fully quantum treatment, based on the explicit description of incoherent fields and the correspondingly induced dipole moments, and of a coherent field with its accompanying polarization, together with an analysis based on nonlinear dynamical properties permits the clear and unequivocal identification of a threshold for the emergence of a self-sustained stimulated emission, i.e., the lasing onset.  The Coherent-Incoherent model marks an entirely new approach in the depiction of laser action, due to the traditional attention brought to macroscopic devices and to the resulting attempts at adjusting the latter to cover small devices through simple modifications.  This treatment shows that simple adjustments are not sufficient and that a consistent treatment can be obtained only through fundamentally revisiting the physics to explicitly introduce the two categories of incoherent and coherent variables.

The main result is a proper definition of lasing threshold irrespective of laser size, accompanied by a continuous description of the evolution of the degree of coherence from the macro- to the nano- scale; we further find that the number of coupled emitters contribute to sharpening or softening the more extreme aspects of the system size.  

In addition to the definition of threshold based on nonlinear physics concepts, the quantum mechanical approach permits the direct evaluation of the coherence properties of the electromagnetic field through the first order coherence function.  Its use shows that full coherence is attained at the bifurcation point (laser threshold), which -- at variance with scaling laws established at the macroscale -- is placed closer and closer to the upper emission branch (or directly on it) as the finite system size contribution increases through the reduced number of electromagnetic cavity modes.  Simultaneously, the quantum-mechanical analysis shows that the rapid growth in photon number originates from an increase in correlation between absorption and emission processes in the absence of self-sustained stimulated emission, which account for the entirety of the transition to the upper emission branch in the smallest devices.  In macroscopic lasers, instead, this contribution is limited to the lower portion of the (nearly) vertical growth in photon number.

A remarkable aspect of the CI model rests on its ability to predict features experimentally observed in measurements of fringe visibility~\cite{tempel11}; a good qualitative agreement is obtained without any free parameters between observations and the predictions shown in this paper.  The topic is of great interest since it allows for an unequivocal quantification -- and for general model-based predictions -- of the amount of coherence, potentially paving the way to numerous applications ranging from novel uses for micro- and nanolasers, but also permitting better assessment of their performance as sources for data treatment (e.g., interconnects in data centers with ultra-low dissipation and small footprint~\cite{miller09,service10,smit12,notomi14,soref18,ning19a}). 

The availability of a complete description of the threshold physics at all scales permits the comparison with other experimental choices.  For instance, one can envisage computing the output of a mixing interferometer~\cite{lebreton13a} to interpret its results on the basis of a first-principle model, rather than superposing \textit{ad hoc} radiation packets with preset features.

The simultaneous availability of first-order and second-order autocorrelations, in addition to the threshold information gathered through the LSA, also permits a careful evaluation of the individual properties of these indicators.  This way, second-order autocorrelation measurements, easier to perform and routinely used not only in Quantum-Dot-based devices~\cite{ulrich07,wiersig09,ota17,Kreinberg2017a}, but also with Quantum-Well emitters~\cite{wang15,takiguchi16,wang20} and metallic nanolasers~\cite{hayenga16,pan16}, can acquire a higher degree of reliability in the determination of the nature of the emitted radiation.  This can contribute to reaching an agreement on a definite measurement technique for the determination of laser threshold~\cite{lasthr} thus sorting the different practical definitions used over the past decades, which give concordant results only at the macroscopic scale~\cite{lippi22}.

The CI models conclusively show that the transition from incoherent to coherent radiation occurs in a negligibly small pump interval for macroscopic devices.  However, they also prove that the physics of laser threshold remains the same even for large lasers, thus implying that the only obstacle in obtaining information from an experiment is of practical nature.  This interpretation is consistent with the results of pioneering work of the 1960's and 70's~\cite{arecchi66,arecchi71}, where statistical ensemble measurements gave evidence for a gradual evolution in the nature of the emitted radiation at threshold crossing.  More information could become now available through the realization of a novel system constituted by a broadband semiconducting amplifier, where feedback is provided by a fiber loop (also containing adjustable filters) which permit stroboscopic measurements of the light amplification as a function of round trip~\cite{roche22}.  The degree of spatio-temporal resolution gained from this realization, thanks to the long delay time of the fibered cavity, enables the measurement of the radiation properties at each round trip.  This scheme could garner detailed information to refine our understanding and mathematical description of laser threshold.

%

\end{document}